# Diamondization of Graphene and Graphene-BN Bilayers: Chemical Functionalization and Electronic Structure Engineering

Long Yuan, Zhenyu Li, Jinlong Yang,* Jian Guo Hou

Hefei National Laboratory of Physical Sciences at Microscale, University of Science and Technology of China, Hefei, Anhui 230026, China

## Abstract

In this article, based on first-principles calculations, we systematically study functionalization induced diamonization of graphene bilayer and graphene-BN hybrid bilayer. With single-side functionalization, the diamondized structures are magnetic semiconductor. Interestingly, if both sides of the bilayer are functionalized, diamondization becomes spontaneous without a barrier. On the other hand, when the bottom layer of the bilayer graphene is replaced by a single hexagonal BN layer, the diamondized structure becomes nonmagnetic metal. The tunable electronic and magnetic properties pave new avenues to construct graphene-based electronics and spintronics devices.

# INTRODUCTION

Graphene has attracted considerable attentions due to its remarkable physical properties such as massless Dirac fermion behavior,[1] room temperature quantum Hall effect,[2] and high carrier mobility.[3-5] It is suggested as a promising electronic material to replace the currently used scilion.[6] Since graphene is semiconducting with a zero energy gap, exploring effective approaches to open a band gap is essential for its potential electronic applications. A finite gap can be opened by using suitable substrates or patterning graphene into nanoribbons.[7-10] Alternatively, chemical functionalization is also an important route to introduce a band gap.[11] For example, fully hydrogenated single-layer graphene, also referred as "graphane", has been predicted theoretically and synthesized experimentally,[12-14] and it is a nonmagnetic semiconductor with a direct band gap of 3.43 eV.[15] A theoretical work based on density functional theory (DFT) shows that single-side half hydrogenated graphene, namely "graphone", is a ferromagnetic semiconductor with a small bandgap.[16] Other functionalized graphene such as graphene fluoride and graphene oxide are also investigated to modify electronic and magnetic properties of pristine single-layer graphene.[17-19]

Compared with single-layer graphene, bilayer graphene shows some unique properties. For example, bilayer graphene can open a gap by inducing an external electric-field or via surface chemical modification.[20-21] Theoretical investigation has predicted that hydrogenated bilayer graphene is a ferromagnetic semiconductor with a tunable band gap.[21] A recent experiment has presented evidence for the room-temperature dimondization of hydroxylated bilayer graphene via increasing pressure. Density functional calculations show that the resulting material, which is called diamondol, is a ferromagnetic insulator, with a band gap of 0.6eV and a magnetic moment of 1 Bohr per unit cell.[22]

Motivated by this interesting experiment, we raise the following questions. Is it possible to functionalize bilayer graphene with other groups to obtain tunable electronic structures? Is it possible to realize dimondization without applying external pressure? Will the electronic structure largely change if we substitute one graphene layer to a BN layer? To answer these questions, first-principles calculations are performed. Our results demonstrate a great flexibility of electronic structure engineering by chemical functionalization.

**THEORY AND METHODS**

All geometry optimizations and electronic structures calculations are based on spin-polarized DFT as implemented in the Vienna *ab initio* Simulation Package (VASP). The project-augmented wave method for the core-valence interaction and the Perdew-Burke-Erzerhof (PBE) generalized gradient approximation (GGA) for the exchange-correlation interaction is employed.[23-26] It is well known that GGA overestimates the equilibrium distance of bilayer graphene. The van der Waals interaction (vdW) correction is employed to obtain a reasonable bilayer distance.[27] A kinetic energy cutoff of 500eV is chosen in the plane-wave expansion. A thick (12 Å) vacuum layer is used to avoid the interaction between two adjacent periodic images. Reciprocal space is represented by Monkhorst-Pack special k-point grid.[28] The Brillouin zone is sampled by a set of 8×8×1 k-points for the geometry optimizations and 21×21×1 k-points for the static total energy and density of states calculations. Structures are relaxed using the conjugate gradient scheme without any symmetry constraints. The convergence of energy and force is within $1\times10^{-4}$ eV and 0.01 eV/Å. The accuracy is first tested by calculating the C-C and B-N bond length of graphene and hexagonal BN. Our results are 1.42 and 1.45 Å, in good agreement with the experimental values. The vdW correction gives a bilayer distance of 3.24 Å, consistent with the experiment value of graphite.[29] Since GGA underestimates band gap, we have also performed some test calculations with the hybrid Heyd-Scuseria-Ernzerhof (HSE) functional.[36-37] Although band gap increases with HSE functional, the main results reported here remain unchanged.

## RESULTS AND DISCUSSION

### Single side functionalization with different groups

For ease of discussion, we name $m$-layer diamondized nanofilm decorated with different chemical groups as X- $C_m$-Y (X, Y= H, OH, and F). The experimentally obtained hydroxylated diamondol is OH-$C_2$. We first investigate the single side decorated bilayer graphene with H and F. They can also form similar structures as diamondol, noted as H-$C_2$ and F-$C_2$. In all optimized structures, C atoms in the first layer become fully $sp^3$-hybridized. In the second layer, only half the carbon atoms become $sp^3$-hybridized, the other C atoms remain $sp^2$-hybridized. The relaxed bond lengths of C-H, C-O and C-F are 1.12, 1.44, and 1.39 Å, respectively. The interlayer C-C bond distance of H-$C_2$, OH-$C_2$, and F-$C_2$ are 1.68, 1.66, and 1.67 Å, respectively.

Interesting magnetic property is obtained for the diamondized structures. As we mentioned, there are still half carbon atoms in the second layer remaining $sp^2$-hybridized, where a localized electron is expected, which gives about 1 $\mu_B$ magnetic moment per unit cell. In order to reveal the preferred coupling of these moments, we calculate total energy for both ferromagnetic (FM) and antiferromagnetic (AFM) configurations using a supercell with four unit cells. The results are listed in table 1, with the FM coupling energetically more favorable. Using mean-field theory, we can estimate the Curie temperature ($T_c$) by the formula $\gamma k_B T_c/2 = E_{AFM}-E_{FM}$, where $\gamma$ is the dimension of the system, $k_B$ is the Boltzmann constant, $E_{FM}$ and $E_{AFM}$ are total energies per unit cell for the FM and AFM configurations.[17] The Curie temperature of graphone is estimated to be 417 and 278 K, when it is treated as 2D and 3D, respectively. As shown in table 1, Curie temperatures of the diamondized structure are higher than $T_c$ of graphone.

In order to visualize the distribution of magnetic moments on the diamondized structure, we plot spin density isosurface in Figure 1. It clearly indicates that the magnetic moments are mainly localized on the sp$^2$-hybridized carbon atoms contributed by the unpaired $p_z$ electrons. To understand more details about their electronic structures, band structures and projected density of states are also plotted in Fig 1. We can see that H-C$_2$ and F-C$_2$ are both indirect gap semiconductor, with a band gap value of 0.54 eV. OH-C$_2$ has a direct band gap of 0.35 eV. Therefore, bilayer graphene can modified with different groups, with tunable electronic structures.

Another important issue is the kinetics of the diamondization. Climbing image nudged elastic band method is employed to obtain the minimum energy path (MEP) and determine the energy barrier.[30-31] Take F-C$_2$ as an example, a supercell consists of four unit cells are computed for the MEP and the caculated barrier is 0.887eV, so there is a barrier of about 0.22 eV per unit cell for diamondization, as shown in Figure 3a. In the initial state (I), when half of the carbon atoms in the top layer are fluorinated, strong σ-bond between C atoms and F atoms is formed and the original extensive π-bonding network of graphene is broken, leaving the electrons in the unfluorinated C atoms localized and unpaired. So the fluorinated top graphene layer exhibits a buckling structure with a thickness of 0.31 Å. The computed equilibrium distance between the two layers is 3.10 Å. In the transition state (T), the thickness of the top and bottom layers is 0.42 and 0.21 Å, respectively. The distance between the two layers is 2.01 Å. In the final state (F), thickness of the top and bottom layers is 0.48 and 0.36 Å, respectively, reaching their maximum value. Interlayer bonds are formed, which gives a diamondized structure. MEPs of other diamondized films, like H-$C_2$ and OH-$C_2$, are also computed. They show similar behavior, with a barrier of around 0.23 eV per unit cell.

**Double-side functionalized bilayer graphene**

**Homogeneous surface decoration.** Now, we study the possibility of double-side functionalization by H, OH, and F. We check the homogeneous surface decoration first. During geometry optimization, three homogeneous diamondized nanofilms are formed, labeled as H-$C_2$-H, OH-$C_2$-OH, and F-$C_2$-F. As shown in Fig 3, now all C atoms become $sp^3$-hybridized. The optimized bond length of C-H, C-O and C-F are 1.11, 1.42, and 1.38 Å, respectively, slightly smaller than the single side decorated systems. The interlayer C-C bond distance of H-$C_2$-H, OH-$C_2$-OH, and F-$C_2$-F are 1.56, 1.55, and 1.55 Å, respectively, about 0.1 Å shorter than the single side decorated film, indicating the formation of stronger chemical bonds.

There is no explicit magnetism in double-side functionalized films. Their band structures are plotted in Fig 4. All three films are found to be direct semiconductors with band gaps of 2.75, 2.85, and 3.64 eV, respectively, much larger than the single side decorated films. The valence-band maximum (VBM) and conduction-band minimum (CBM) are both located at the Γ point in the reciprocal space.

More interestingly, the diamondization process is different in the double-side functionalized case. Taking F-$C_2$-F nanofilm as an example, its minimum energy path for one unit cell is shown in Fig 3b. The transformation is proved to be a barrierless process. Similar barrierless process is also observed in other double-side functionalized films, such as H-$C_2$-H and OH-$C_2$-OH. Therefore, extra pressure may not be required to synthesize these double-side functionalized films.

**Heterogenous surface decoration.** We also investigate the bilayer graphene decorated with H and F, H and OH, and F and OH on each side. Similar diamondized nanofilms, noted as H-$C_2$-F, H-$C_2$-OH, and F-$C_2$-OH, are obtained, as shown in Fig 5. All the C atoms in the heterogenous functionalized films are sp$^3$-hybridized and show no magnetic moment. The relaxed bond length of C-H, C-O and C-F are 1.11, 1.42, and 1.38 Å, respectively. The interlayer C-C bond distance of H-C$_2$-OH, H-C$_2$-F, and F-C$_2$-OH is 1.55, 1.56, and 1.55 Å, respectively, almost the same as in the homogeneous functionalized case. The calculated band structures are also shown in Fig 5. All three films are semiconducting with direct band gaps. The computed energy gaps of H-C$_2$-OH, H-C$_2$-F, and F-C$_2$-OH are 1.42, 2.23, and 2.97eV, respectively.

**Diamondized graphene/ *h*-BN bilayer**

Recently, graphene supported on a hexagonal boron nitride (*h*-BN) substrate has attracted remarkable attention. Graphene supported on *h*-BN substrate has a band gap and exhibits higher mobility compared to other substrates.[32-34] Functionalization of h-BN hybridized with graphene has also been studied theoretically.[35] Here, we consider the Bernal stacking of graphene on *h*-BN substrate, where half of the C atoms in the graphene are positioned exactly above the B atoms. This stacking has been proved to be the lowest-energy configuration.[32] When the graphene layer is decorated with different chemical groups (the number of each kind of functional groups equals to half of the number of C atoms), new hybrid diamondized films are obtained, labeled as H-C-BN, OH-C-BN, and F-C-BN, as shown in Fig 6. The optimized bond length of C-H, C-O and C-F are 1.12, 1.47, and 1.41 Å, respectively. The interlayer C-B bond length is 1.92, 1.80, and 1.83 Å. Band structures and density of states of H-C-BN, OH-C-BN, and F-C-BN are also presented in Fig 6. Interestingly, all three hybrid films are predicted to be non-magnetic metals. From first sight, this result is surprising. Since BN single layer is an insulator itself, it is not expected that X-$C_2$ will becoming metallic by replacing the lower graphene layer with BN layer. Examination of the states near the Fermi level shows that the N $2p$ and C $2p$ states contributes to the band crossing the Fermi level. Therefore, electron transfer from BN to the top graphene layer is the origin of the metallicity.

**Stabilities of diamondized nanofilms**

To investigate the structural stability of the diamondized nanofilms, formation energy is defined as

$$E_f = (E_{dia} - E_{bilayer} - n\mu_x)/n$$

Where $E_{dia}$ is the total energy per unit cell of diamondized nanofilms, X represents different functional groups (H, OH, and F), $n$ is the number of functional groups per unit cell, $E_{bilayer}$ is the toal energy of bilayer graphene or graphene/$h$-BN (Bernal stacking). We choose $\mu_H$ as 1/2 of the cohesive energy of the $H_2$ molecule, $\mu_F$ as the 1/2 cohesive energy of $F_2$, $\mu_{OH}$ as the 1/2 cohesive energy of the $H_2$ molecule plus 1/2 cohesive energy of the $O_2$ molecule. The stability of different diamondized nanofilms are examined by formation energies, those with smaller formation energies are energetically more stable. As shown in Fig 7, generally hydroxylated and fluorinated films have nearly the same formation energies, much lower than hydrogenated one, indicating that the former two are more easily synthesized in experiments with $H_2$ and $O_2$ provided. At the same time, double-side functionalized films (homogeneous or heterogenous) are more stable than single-side decorated films.

## CONCLUSION

In summary, based on first-principles calculations, we have systematically investigated functionalized bilayer graphene and graphene/*h*-BN. Upon functionalization, they can have a diamondization transition. By using different functional groups, we can tune the electronic structure of the diamondized films. Diamonziation may be spontaneous if double-side functionalization is applied. Interestingly, if one graphene layer is substituted by a BN layer, the system becomes metallic, which is a result of the interlayer charge transfer. Our results demonstrate the flexibility of chemical functionalization on electronic structure engineering of 2D nanostructures, and will inspire experimentalists to synthesize such interesting 2D materials, since all structures proposed here are able to be obtained straightforward experimentally.

**Acknowledgements.** This work is partially supported by NSFC (21121003, 91021004, and 20933006), by MOST (2011CB921404), and by USTC-SCC, SCCAS, and Shanghai Supercomputer Centers.

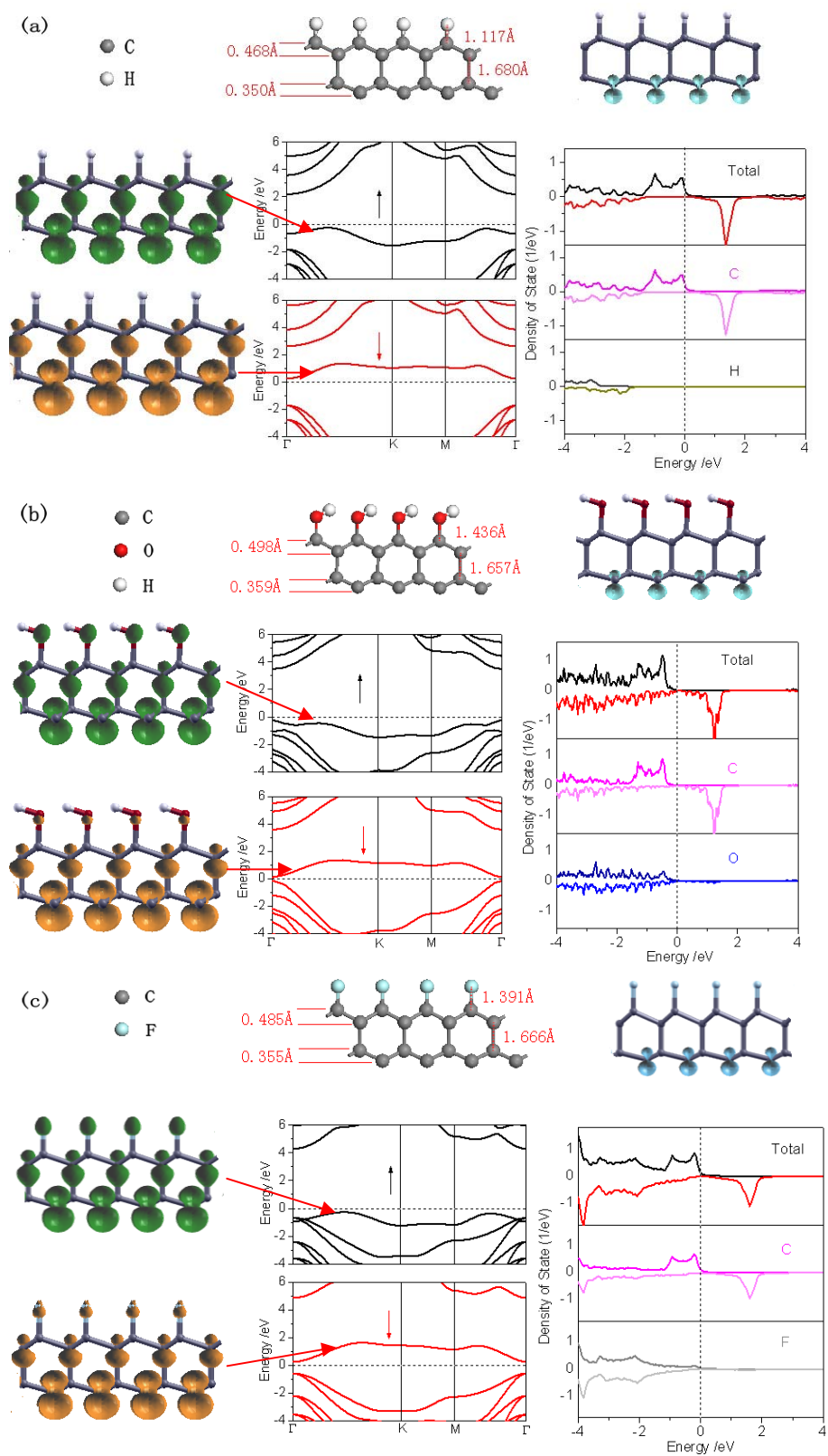

Fig 1. Optimized geometry, spin density distribution (isosurface value:0.08 e/Å$^3$), band structure, and projected density of states of (a) H-C$_2$, (b) OH-C$_2$, and (c) F-C$_2$. Band resolved charge density at left panels are plotted an isosurface value of 0.02 e/Å$^3$.

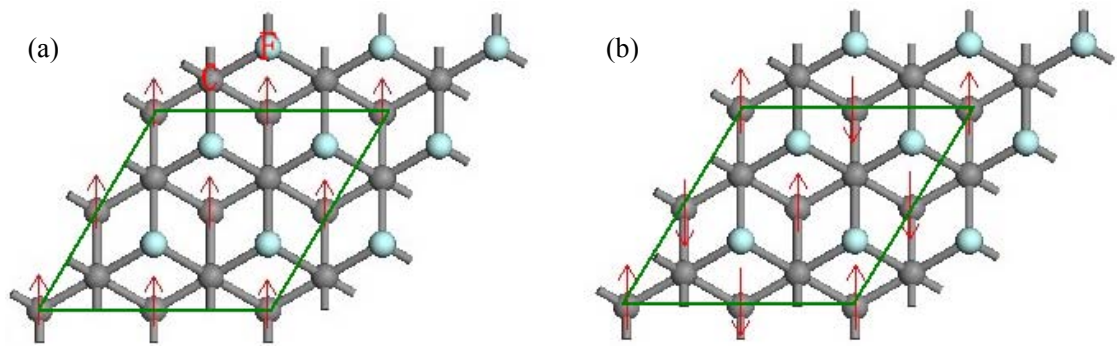

Fig 2. (a) Ferromagnetic and (b) antiferromagnetic configurations of single-side decorated diamondol nanofilms

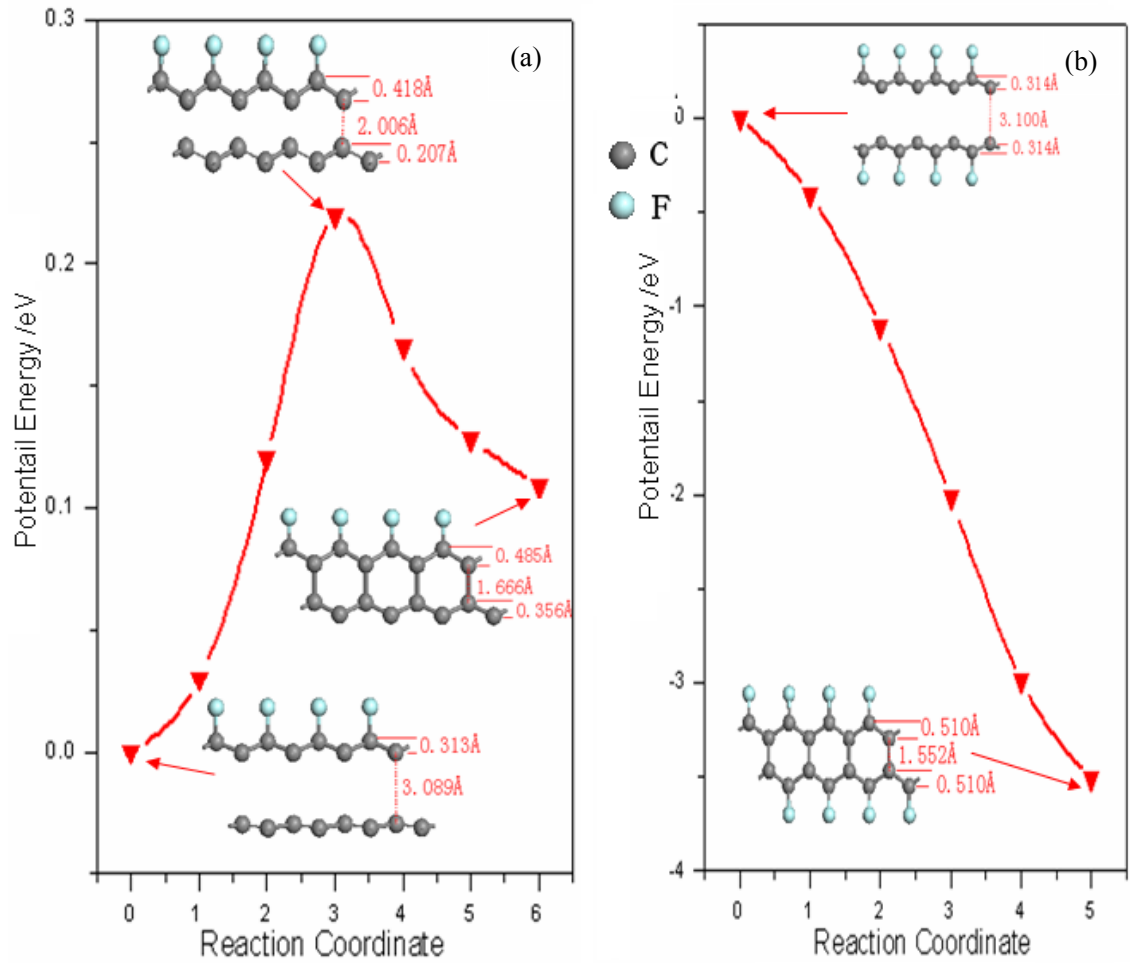

Fig 3. Minimum energy path for transition processes (a) from single-side functionalized bilayer to diamondized film, (b) from double-side functionalized bilayer to diamondized film.

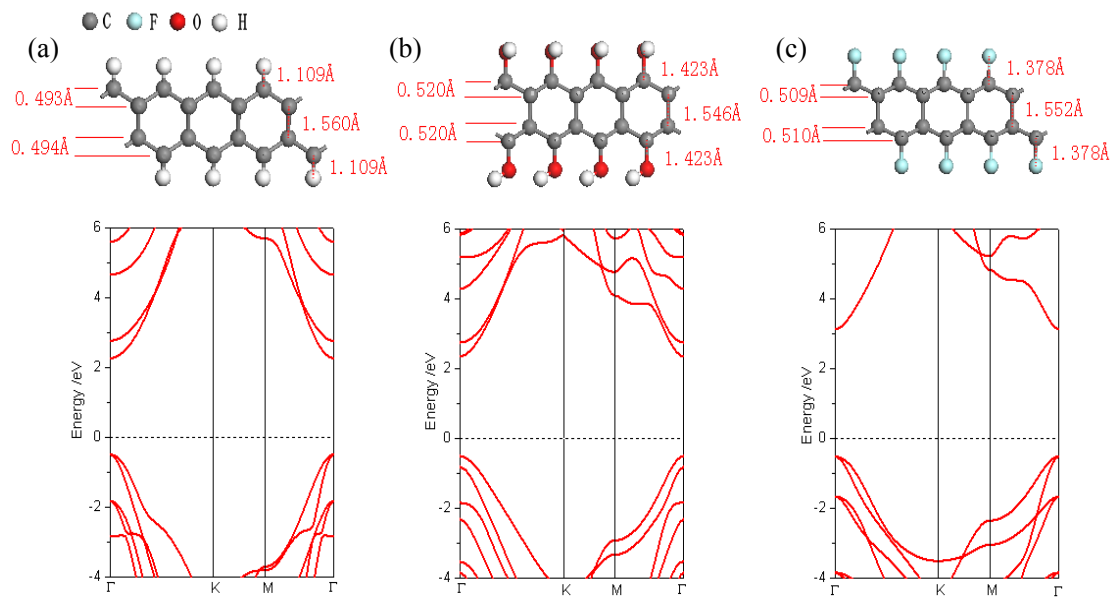

Fig 4. Optizimed geometry and band structure of (a) H-$C_2$-H, (b) OH-$C_2$-OH, and (c) F-$C_2$-F.

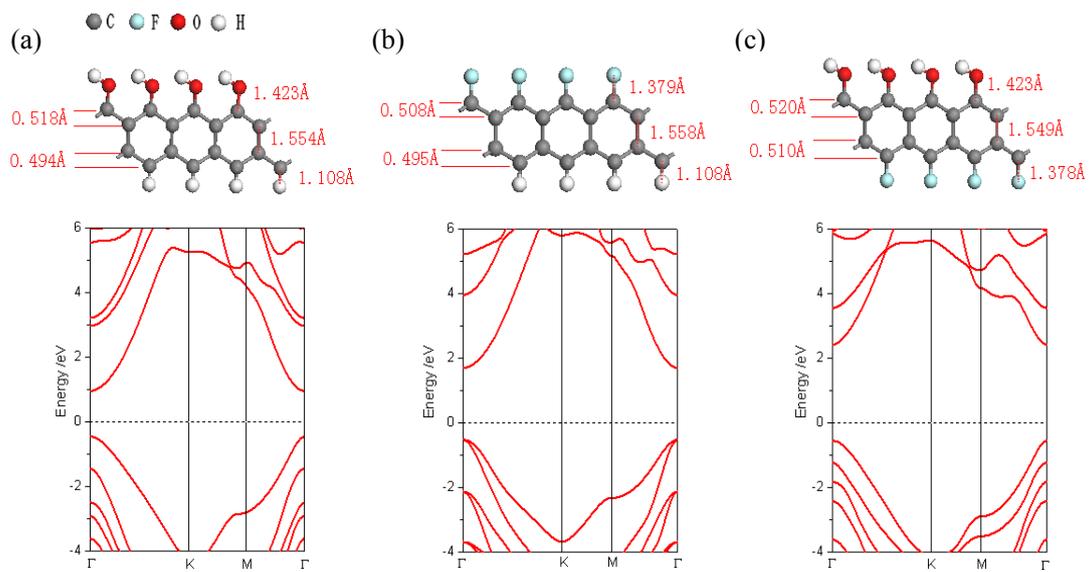

Fig 5. Optimized geometry and band structure of (a) OH-C$_2$-H, (b) F-C$_2$-H, and (c) OH-C$_2$-F.

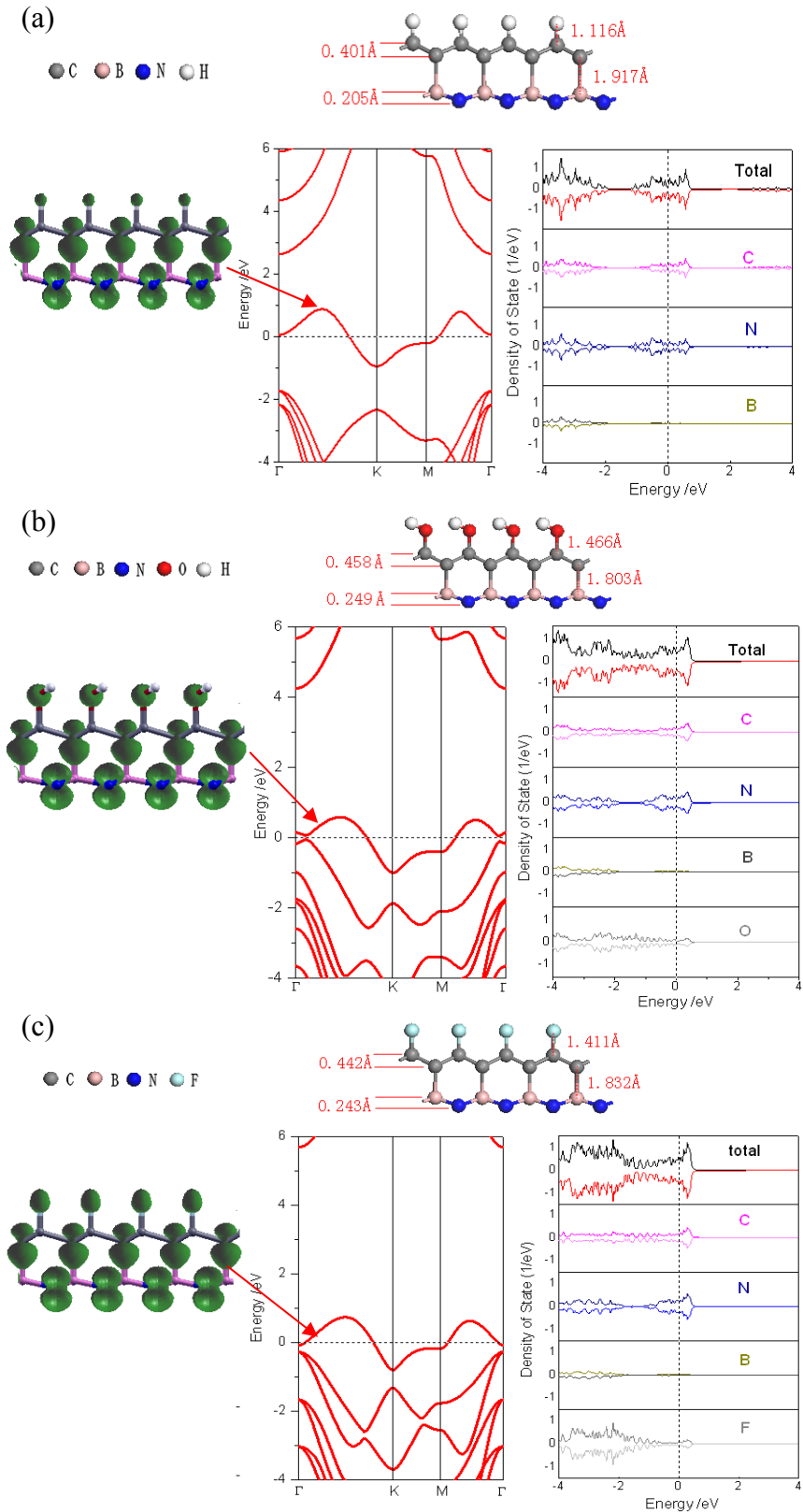

Fig 6. Optimized geometry, band structure, and projected density of states of (a) H-C-BN, (b) OH-C-BN, and (c) F-C-BN. The charge density contributions are plotted with an isosurface value of 0.02 e/Å$^3$.

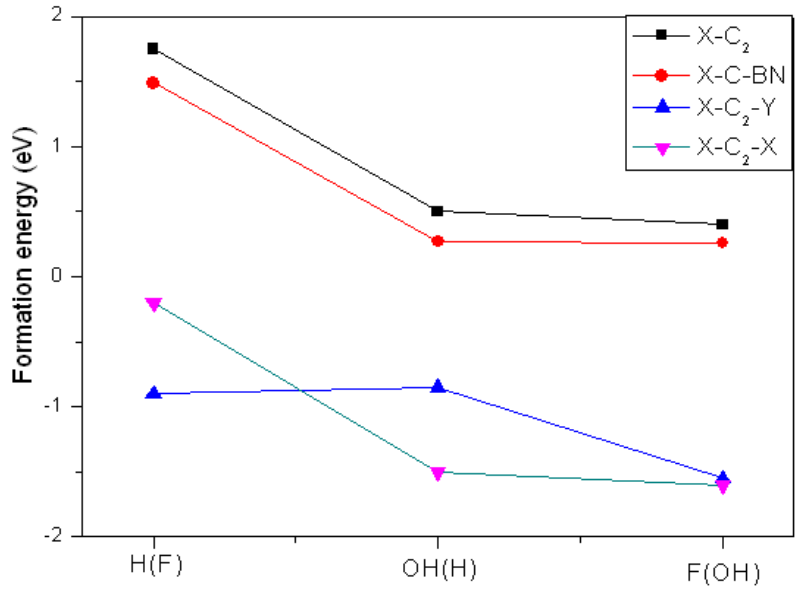

Fig 7. Formation energies of diamondized nanofilms with different functional groups. X = H, OH, F (Y= H, OH, F), respectively.

Table 1. Relative energy per unit cell of ferromagnetic state ($E_{\text{FM}}$) and antiferromagnetic state ($E_{\text{AFM}}$), and Curie temperature ($T_c$) estimated with 2D and 3D models.

|  | $E_{\text{FM}}$ (eV) | $E_{\text{AFM}}$ (eV) | $T_c$(2D) (K) | $T_c$(3D) (K) |
| --- | --- | --- | --- | --- |
| H-$C_2$ | 0.00 | 0.049 | 379 | 569 |
| HO-$C_2$ | 0.00 | 0.047 | 363 | 544 |
| F-$C_2$ | 0.00 | 0.044 | 340 | 510 |

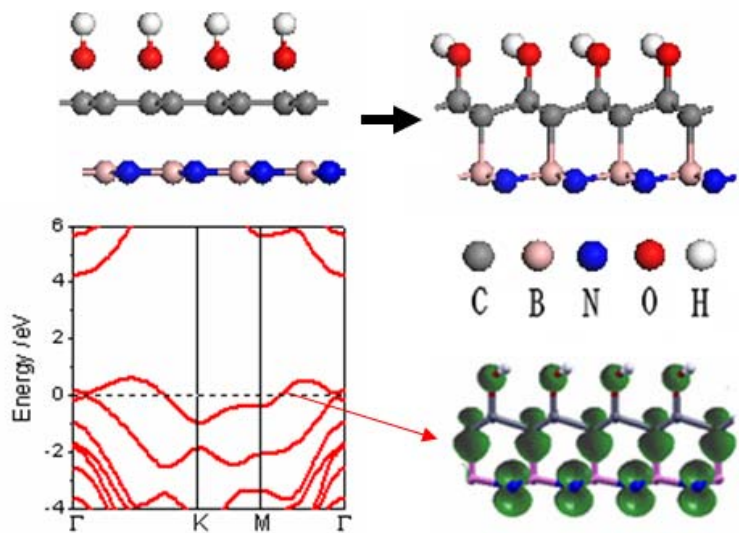

TOC Figure